\documentclass[a4paper]{article}
\pdfoutput=1

\usepackage{todonotes}
\usepackage{lineno}
\usepackage[hidelinks]{hyperref} 
\usepackage[english]{babel}
\usepackage[utf8]{inputenc}
\usepackage{natbib}
\usepackage{amsmath}
\usepackage{amsfonts}
\usepackage{amsthm}
\usepackage{mathtools}
\usepackage{nicefrac}
\usepackage{graphicx}
\usepackage{url}
\usepackage{listings}
\usepackage{array}
\usepackage{verbatim}
\usepackage{longtable}
\usepackage[official]{eurosym}
\usepackage[ruled,linesnumbered]{algorithm2e}
\usepackage{tikz}
\usepackage{booktabs}
\usetikzlibrary{shapes,arrows, shapes.geometric, shapes.symbols}
\everymath{\displaystyle}
\usepackage{authblk}

\DeclareMathOperator*{\argmin}{arg\,min}
\newtheorem{theorem}{Theorem}
\newtheorem{lemma}{Lemma}
\newtheorem{remark}{Remark}
\newtheorem{definition}{Definition}
\renewcommand*{\eqref}[1]{Equation~(\ref{#1})}

\tikzstyle{block} = [rectangle, draw, fill=blue!20, 
    text width=3cm, text centered, rounded corners, minimum height=4em]
\tikzstyle{line} =  [draw, thick, ->, shorten >=2pt] 
\tikzstyle{cloud} = [draw, ellipse,fill=red!20, node distance=3cm,
    minimum height=2em]
\tikzstyle{print} = [draw, tape, tape bend top=none, fill=blue!20, node distance=3cm
	 text width=5em, minimum height=4em, text centered]
\tikzstyle{decision}= [diamond, aspect=2, draw, fill=blue!20,
                        text width=2.5cm, text centered,
                        inner sep=1pt, rounded corners]

\usepackage{pgfplots}
\usepgfplotslibrary{fillbetween}
\usetikzlibrary{calc}
\usetikzlibrary{patterns}
\usepgflibrary{patterns}
\usepgfplotslibrary{statistics}
\pgfplotsset{compat=1.16}

\usepackage{tabularx,booktabs,siunitx}

\usepackage{glossaries}
\glsdisablehyper

\newacronym{iqr}{IQR}{\emph{Interquartile Range}}
\newacronym{gdm}{GDM}{\emph{Group Decision-Making}}

\begin{document}

\title{A General Approach for Computing a Consensus\\in Group Decision Making\\That Integrates Multiple Ethical Principles}

\author[1]{Francisco Salas-Molina\footnote{Corresponding author. E-mail addresses: \textit{frasamo@upv.es}}}

\author[2]{Filippo Bistaffa}

\author[2]{Juan A. Rodríguez-Aguilar}

\affil[1]{Universitat Polit\`enica de Val\`encia, Ferrandiz y Carbonell s/n, 03801 Alcoy, Spain}

\affil[2]{IIIA-CSIC, Campus UAB, 08913 Bellaterra, Spain}

\maketitle

\begin{abstract}
We tackle the problem of computing a consensus according to multiple \emph{ethical principles} ---which can include, for example, the principle of maximum freedom associated with the Benthamite doctrine and the principle of maximum fairness associated with the Rawlsian principles--- among the preferences of different individuals in the context of \gls{gdm}. More formally, we put forward a novel formalisation of the above-mentioned problem based on a \emph{multi-$\ell_p$-norm approximation problem} that aims at minimising multiple $p$-metric distance functions, where each parameter $p$ represents a given ethical principle. Our contribution incurs obvious benefits from a social-choice perspective. Firstly, our approach significantly generalises state-of-the-art approaches that were limited to only two ethical principles ($p=1$, for maximum freedom, and $p=\infty$, for maximum fairness). Secondly, our experimental results considering an established test case demonstrate that our approach is capable, thanks to a novel re-weighting scheme, to compute a multi-norm consensus that takes into account each ethical principle in a balanced way, in contrast with state-of-the-art approaches that were heavily biased towards the $p=1$ ethical principle.
\\

\\
\end{abstract}


\glsreset{gdm}

\section{Introduction}

\gls{gdm} is ubiquitous in everyday life.
Even when decision-makers act individually, they often receive advice or suggestions from others.
Thus, decisions are often social in nature and involve multiple group members \citep{tindale2019group,yazdani2022fuzzy}.
Among others, one prominent example of \gls{gdm} is \emph{citizen participation} \citep{fagence2014citizen}, adopted by the public administrations of many major cities all over the world. As an example, the citizen participation platform by the Barcelona administration\footnote{Accessible at \url{https://ajuntament.barcelona.cat/participaciociutadana}.} is actively being used to decide and implement local policies with the direct involvement of citizens. 
Along these lines, reaching an agreement that represents a \emph{principled} compromise among multiple preferences or opinions is of utmost importance in modern society.

Over the years, a vast wealth of literature concerned with \gls{gdm} has formulated and solved this problem in many different ways 
\citep{butler2006conflict,xu2011group,herrera2014review,ortega2015modelling,perez2018dynamic,ZHANG202071,LU2021910,noori2021group,pamucar2022integrated,ZHANG2022340,LIU2022269,LIANG2022361,HASSANI202286,zhang2021social,xiao2022exploring}.
In this paper, we follow the approach introduced by  \cite{gonzalez1999distance,gonzalez2008aggregation,gonzalez2011design,gonzalez2016bentham}, where the problem of finding a consensus or a collective decision is achieved by minimising a $p$-metric distance function among the preferences (or opinions) of the individuals.

One major benefit of this approach is that it allows decision-makers to compute a consensus following different \emph{ethical principles} according to different values of the parameter $p$.
For instance, $p=1$ corresponds to the idea of maximum freedom of the individuals derived from the theory of utilitarianism by \cite{bentham1789introduction}, whereas $p=\infty$ corresponds to the principle of maximum fairness, from the Rawlsian idea of considering only the welfare of the worst-off group \citep{rawls1973theory}.\footnote{In this paper, we use ``norm'', ``metric'' and ``ethical principle'' to refer to the same concept.} The principle of maximum freedom is relevant when the increase in the welfare of a well-off group by one unit has the same social value as an increase in the welfare of a disadvantaged group by one unit. On the contrary, in the principle of maximum fairness, the welfare of society only depends on the utility of the worst-off individual or social group. 
For instance, let us consider the problem faced by policy-makers in deciding whether to construct a new parking lot or a bike lane.
Assuming that the majority of the population use cars as a personal means of transportation, according to the maximum freedom principle the construction of a new parking lot yields the maximum welfare for the society.
On the other hand, the worst-off group (i.e., cyclists) would benefit from the construction of a new bike lane, hence it would adhere to the maximum fairness principle.

Apart from these two well-known cases, an exciting research question worthy of investigation is how to compute a consensus for any of the infinite values of $p\in\mathbb{N}$ that represent \emph{intermediate} ethical principles other than maximum freedom or maximum fairness.
Unfortunately, the approach proposed by 
\cite{gonzalez2016bentham} cannot be applied to values of $p$ different from $1$ and $\infty$, due to the computational limitations of the goal programming solution technique adopted by the authors.

Besides that, and taking a more general perspective, computing a consensus may require the involvement of \emph{multiple} ethical principles, be it because the decision-making involves multiple stakeholders (each one with their own ethical principle), or because the consensus aims at a trade-off between ethical principles. 
To the best of our knowledge, despite previous attempts considering ad-hoc combinations of $p=1$ and $p=\infty$ \citep{gonzalez2016bentham}, computing a consensus that involves a general combination of multiple, any-valued ethical principles is still an open problem that has not been addressed by the literature.

Against this background, in this paper we propose a novel generalisation to the previous approach by \cite{gonzalez1999distance,gonzalez2008aggregation,gonzalez2011design,gonzalez2016bentham} based on an \emph{$\ell_p$-norm approximation problem} \citep{boyd2004convex} (also called \emph{$\ell_p$-regression} \citep{adil2019fast}), which is able to account for multiple, any-valued ethical principles.
Thus, we show that computing a consensus for a multi-principle problem amounts to solving a  multi-$\ell_p$-norm approximation.

Our generalisation is conceived to provide expressiveness for: (i) each stakeholder to choose their own ethical principle; and (ii) setting the relative importance of ethical principles, and hence the preferences over ethical principles (e.g., three different groups supporting three ethical principles setting the relative importance by means of the size of the group). Such a novel approach yields multiple benefits both from the theoretical and the computational perspectives. Most importantly, our approach can be used to compute a consensus for a given set of ethical principles, and also in particular for any value of $p$ (not only for $p=1$ and $p=\infty$), hence representing a significant generalisation of state-of-the-art work on the matter.

Importantly, our approach has been conceived to pursue a \emph{balanced} contribution of each ethical principle to the consensus. This is achieved by carefully aggregating the multiple $p$-norms involved in the computation of a consensus. More precisely, we introduce a \emph{re-weighting approach} for multi-$\ell_p$-norm approximation that aims at \emph{normalising} the values of the different $p$-norms involved prior to their aggregation. Finally, following the methodology proposed in \cite{gonzalez2016bentham}, we empirically confirm that our re-weighting approach is indeed necessary to fairly include the contributions from the different ethical principles at hand.

In more detail, this paper advances the state-of-the-art in the following ways:
\begin{enumerate}
    \item We show that the problem of computing the consensus among individuals under multiple $p$-metric distance functions \citep{gonzalez2016bentham} can be cast as a \emph{multi-norm approximation problem}, a particular and novel type of norm approximation problem \citep{boyd2004convex}. We discuss how this enables the use of convex optimisation solution techniques as the means to compute the consensus for \emph{multiple} ethical principles.
    \item By doing so, our multi-$\ell_p$-norm approximation approach provides a general theoretical framework that can accommodate \emph{many} values of $p$ (i.e., many ethical principles), hence generalising the approach in \cite{gonzalez2016bentham}. Furthermore, our approach also allows us to compute the consensus for \emph{any} single value of $p$ (in contrast with \cite{gonzalez2016bentham}, which can only deal with $p=1$ and $p=\infty$).
    \item We also propose a \emph{re-weighting approach} for our multi-$\ell_p$-norm approximation (henceforth referred to as \emph{weighted model}) so that the ethical principles involved fairly contribute to the consensus, improving upon the approach proposed in \cite{gonzalez2016bentham}. This is achieved by \emph{normalising} the values of the different $p$-norms involved in computing the consensus. 
    \item Considering an established test case, we illustrate our general methodology for single ethical principles (besides $p=1$ and $p=\infty$) by analysing the distribution of the differences in individual preferences with respect to the consensus for several values of $p$. We exploit this new capability provided our general approach to empirically analyse the consensus for the whole space of intermediate values of~$p$, highlighting that values different from $p=1$, $p=2$ and $p=\infty$ may have a complex semantic interpretation.
    \item Considering the same test case, we empirically analyse how considering different combinations of multiple ethical principles in our multi-$\ell_p$-norm approximation model impacts the consensus, both for the unweighted model (without re-weighting) and for the weighted one (with re-weighting). We observe that using an unweighted model would lead to a biased consensus since the contribution of the utilitarian principle ($p=1$) dominates the contributions of other ethical principles. Nonetheless, we also observe that our weighted model manages to correct such bias. Thus, this model leads to a balanced contribution of ethical principles to the consensus. 
\end{enumerate}


The structure of this paper is as follows. Section \ref{sec:review} reviews related works. Section \ref{sec:back} provides useful background about the use of distance functions in social choice theory. Section \ref{sec:multinorm} describes our novel $\ell_p$-norm approximation approach to account for multiple ethical principles, the so-called \emph{multi-$\ell_p$-norm approximation}. Section \ref{sec:weights} introduces a \emph{re-weighting approach} for multi-$\ell_p$-norm approximation that aims at ensuring the balanced contribution of multiple p-norms to the consensus. Section \ref{sec:exp} empirically analyses the semantics of single ethical principles and analyses the impact of considering multiple ethical principles on a consensus.Section \ref{sec:conc} offers some concluding remarks and a natural extension of this work.

\section{Literature review\label{sec:review}}

The problem of computing a consensus or an aggregation of preferences of different individuals (or within social groups) has been widely studied in the literature, which usually considers individuals that have to express their opinions on some alternatives or policies \citep{butler2006conflict}.
In many cases, preferences over alternatives are expressed by preference relations \citep{xu2011group}.
In this context, authors have considered a variety of different settings, including dynamic scenarios \citep{perez2018dynamic,HASSANI202286}, scenarios with fuzzy preferences \citep{herrera2014review,ZHANG202071,ZHANG2022340} or characterised by a social network structure \citep{LU2021910,LIU2022269,LIANG2022361}. Some authors study the impact of the social trust relationships on assessments-modifications in the consensus reaching in social network group decision making \cite{zhang2021social}. Other authors describe a multi-stage consensus optimisation model with bounded confidence to obtain the adjustment suggestions by maximising the level of consensus \cite{xiao2022exploring}. If the predetermined level of consensus cannot be reached, the adjustment suggestions obtained by the model are adopted to guide the preference modification of the group members.

On the one hand, a body of literature has focused on the specific case where each individual expresses their preferences in the form of a \emph{ranking}. Finding this optimal aggregated ranking can be formulated as an optimisation problem, which is usually NP-hard \citep{dwork2001rank}.
Thus, several heuristic \emph{ranking aggregation methods} have been proposed, including batch mode methods, instant-runoff mode methods, and, more recently, hierarchical methods \citep{ding2018new,aledo2021highly}. Other related works propose methods to overcome the difficulties of eliciting preferences over too many alternatives including a branch-and-bound algorithm to construct a consensus ranking \citep{cook2010aggregating}, an efficient method for aggregating measurements acquired by an uncalibrated environmental sensory network \citep{fishbain2016self}, an axiomatic method to aggregate a set of incomplete rankings into a consensus ranking \citep{moreno2016axiomatic}, a binary programming formulation for the generalised Kemeny \citep{kemeny1962preference} rank aggregation problem \citep{yoo2021new}, and a multimodal data aggregation methodology for jointly aggregating heterogeneous ordinal and cardinal evaluation inputs into a consensus evaluation \citep{escobedo2022axiomatic}.

On the other hand, in some situations it might be difficult or even impossible for individuals to express their preferences as rankings, hence the above-discussed methods cannot be applied.
Indeed, in this paper, we follow an alternative and more general approach proposed in another strand of literature \citep{gonzalez1999distance,gonzalez2008aggregation,gonzalez2011design,gonzalez2016bentham} where preferences are expressed by a \emph{pairwise representation}, i.e., a square matrix $R$ where the value $R_{i,j}$ represents the preference of object $i$ with respect to object $j$ (see Section \ref{sec:back} for more details). However, we should mention that pairwise comparison matrices require $m^2$ inputs from each individual whereas a ranking requires only $m$ inputs.

In this context, the problem of finding a consensus between individuals (or within social groups) can be approached by means of a $p$-metric distance function that allows consideration of several ethical principles by varying the value of $p\in\mathbb{R}$ belonging to the closed interval $\left[1, \infty \right]$ as proposed by \cite{gonzalez1999distance,gonzalez2008aggregation,gonzalez2011design,gonzalez2016bentham}. Specifically,  \cite{gonzalez2011design,gonzalez2016bentham} consider particular values of $p$,  evaluating the quality of consensus among groups by the aggregated value of the distance between the position of each group and the consensus point.
To solve the problem, the authors relied on extended goal programming \citep{romero2001extended,jones2010practical}, which can only be used in the case of particular values of $p$ (i.e., $p=1$ and $p=\infty$), as using such a technique for $p > 1$ is computationally too demanding.

Finally, in the context of measurement of economic inequality, \citep{atkinson1970measurement} proposed the use of a parametric inequality measure in which a parameter in the range $\left[1,\infty \right]$ plays the role of a degree of inequality aversion or the relative sensitivity to transfers at different income levels. Instead, in this paper we use $p$ as a degree of utilitarian aversion.

In what follows, we provide detailed background on the social choice problem based on parametric distance functions following the notation used in \cite{gonzalez2016bentham}.

\section{Parametric distance functions in social choice theory\label{sec:back}}

In this section, we describe the social choice problem based on $p$-metric distance functions following the notation used in \cite{gonzalez2016bentham}. We consider a society with $n$ individuals indexed by $i=1, 2, \ldots, n$ that has to provide judgements on $m$ objects or alternatives indexed by $j,k=1,2, \ldots, m$. These judgements can be cardinal, when the individuals express some degree of preference over the alternatives, and ordinal when they use a Boolean value to express their preference for one alternative over another one. The rest of the elements of the problem are as follows:

\begin{itemize}
    \item Weight $w_i$ represents the social influence of the $i$th individual (or social group).
    \item Judgement $R_{jk}^i$ is the value provided by the $i$th individual when comparing the $j$th and the $k$th object.
    \item Judgement $R_{jk}^S$ is the value assigned by the society as a whole when comparing the $j$th and the $k$th object. These are the unknowns, the solution that we seek.
    \item Set $\textbf{F}$ includes the constraints that solutions $R_{jk}^S$ must satisfy.
\end{itemize}

From the previous definitions, the weighted Minkowski $p$-metric distance function ($U_p$) can be used as a generator of social choice functions as described in \cite{gonzalez1999distance, gonzalez2008aggregation,gonzalez2011design,gonzalez2016bentham}. The ultimate goal is obtaining social consensus points $R_{jk}^S$ such that the weighted $p$-metric distance between $R_{jk}^S$ and judgement values $R_{jk}^i$ provided by the individuals within the society is minimised. Here, $p$ is a real number in the closed interval $\left[1, \infty \right]$, which is a key parameter in the following distance function:
\begin{equation}
    U_p = -\left[  \sum_{i=1}^n \sum_{j=1}^m \sum_{k=1, k \neq j}^m w_i^p |R_{jk}^i-R_{jk}^S |^p\right]^{1/p}.
    \label{eq:Mink}
\end{equation}

From this distance function, \cite{gonzalez2016bentham} derived three particular cases to represent well-known ethical principles. By setting $p=1$, the utilitarian principle of freedom to maximise the total welfare $U_B$ proposed by \cite{bentham1789introduction} is represented as:
\begin{equation}
    U_B = -\left[ \sum_{i=1}^n \sum_{j=1}^m \sum_{k=1, k \neq j}^m w_i |R_{jk}^i-R_{jk}^S |\right].
    \label{eq:Bentham}
\end{equation}

By setting $p=\infty$, the idea of fairness is represented by the min-max principle proposed by \cite{rawls1973theory}. This principle, denoted by $U_R$, implies that the maximum deviation from an individual judgement $R_{jk}^i$ with respect to the consensus point $R_{jk}^S$ is minimised:
\begin{equation}
    U_R = - \text{max} \left[ w_i |R_{jk}^i-R_{jk}^S |\right].
    \label{eq:Rawls}
\end{equation}

Finally, the idea of maximum equity by Marx is derived from $U_B$ in \eqref{eq:Bentham}, when considering the following constraint:
\begin{equation}
    w_1 |R_{jk}^1-R_{jk}^S| = w_2 |R_{jk}^2-R_{jk}^S| = \ldots = w_n |R_{jk}^n-R_{jk}^S|.
    \label{eq:Marxeq}
\end{equation}

To address the computational problems of the above models, \cite{gonzalez2016bentham} proposed the following change in variables derived from an initial proposal by \cite{charnes1977goal}:
\begin{equation}
    n_{jk}^i = \frac{1}{2}\big[|R_{jk}^i-R_{jk}^S|+(R_{jk}^i-R_{jk}^S) \big]
    \label{eq:change1}
\end{equation}
\begin{equation}
    p_{jk}^i = \frac{1}{2}\big[|R_{jk}^i-R_{jk}^S|-(R_{jk}^i-R_{jk}^S) \big].
     \label{eq:change2}
\end{equation}

Then, by adding Equations (\ref{eq:change1}) and (\ref{eq:change2}), and by subtracting \eqref{eq:change2} from \eqref{eq:change1}, the following identities are obtained:
\begin{equation}
    n_{jk}^i + p_{jk}^i = |R_{jk}^i-R_{jk}^S|
    \label{eq:change3}
\end{equation}
\begin{equation}
    n_{jk}^i - p_{jk}^i = R_{jk}^i-R_{jk}^S.
     \label{eq:change4}
\end{equation}

By using \eqref{eq:change3} and \eqref{eq:change4}, objective function in \eqref{eq:Mink} becomes:
\begin{equation}
    U_p = -\left[  \sum_{i=1}^n \sum_{j=1}^m \sum_{k=1, k \neq j}^m w_i^p (n_{jk}^i-p_{jk}^i)^p\right]^{1/p}
    \label{eq:Mink2}
\end{equation}
subject to \eqref{eq:change4} and $R_{jk}^S \in \textbf{F}$. As a result, the Benthamite solution $U_B=U_1$ is obtained when $p=1$. To obtain the Rawlsian solution $U_R$, the following goal programming Chebyshev model is solved to obtain $D$, which represents the disagreement of the member of the society with the opinions most displaced from the solution obtained:
\begin{equation}
    U_R = - D
\end{equation}
subject to \eqref{eq:change4}, $R_{jk}^S \in \textbf{F}$, and:
\begin{equation}
    w_i \sum_{j=1}^m \sum_{k=1, k \neq j}^m (n_{jk}^i-p_{jk}^i) -D \leq 0.
\end{equation}

Finally, although it is likely that the system of equations in \eqref{eq:Marxeq} has no solution, the idea of maximum equity ($U_M$) can be approximated by relying on meta-goal programming, as described in \cite{gonzalez2016bentham} and \cite{caballero2006interactive}: 

\begin{equation}
    U_M = - \sum_{i=1}^{n-1}  \sum_{t=i+1}^{n} (\eta_{it}+\rho_{it})
    \label{eq:MarxProg}
\end{equation}
subject to:
\begin{gather*}
    w_i(n_{jk}^i+p_{jk}^i)- w_t(n_{jk}^t+p_{jk}^t)+\eta_{it}-\rho_{it} = 0 \hspace{3mm}
    i = 1, \ldots, n-1 \hspace{3mm}
    t = i +1, \ldots, n \\
    R_{jk}^S+n_{jk}^t-p_{jk}^t =  R_{jk}^i \hspace{3mm}
    i = 1, \ldots, n \hspace{2mm} \forall j,k \\
    R_{jk}^S \in \textbf{F}.
\end{gather*}
where $\eta_{it}$ and $\rho_{it}$ are the usual deviation variables used in goal programming. 

From these particular cases, a convex combination of $U_B$, $U_R$ and $U_M$ can be derived to account for three different ethical principles by varying control parameters $\lambda_1$ and $\lambda_2$ in the interval $\left[0, 1 \right]$:
\begin{equation}
    U(\lambda_1, \lambda_2) = \lambda_1 U_B +\lambda_2 U_R + (1-\lambda_1-\lambda_2) U_M,
    \label{eq:convex1}
\end{equation}
subject to the set of restrictions in $\textbf{F}$ and $\lambda_1+\lambda_2 \in \left[0, 1 \right]$. Note that for $\lambda_1=1$ and $\lambda_2=0$, we obtain the Benthamite solution of maximum freedom; for $\lambda_1=0$ and $\lambda_2=1$, we obtain the Rawlsian solution of maximum fairness; and for $\lambda_1=0$ and $\lambda_2=0$, we obtain the Marxian solution of maximum equity.

As for the set $\textbf{F}$ of constraints that the solution $R_{jk}^S$ must satisfy, three sets of restrictions were considered by \cite{gonzalez2008aggregation}, allowing the solution to reflect any valid set of cardinal and/or ordinal preferences:
\begin{enumerate}
    \item Information about preferences is ordinal and complete. In this case, alternatives are ordered from the best to the worst by using a linear order such that $r_{jk}^i=1$ and $r_{kj}^i=0$ if the $i$th decision-maker prefers the $j$th alternative to the $k$th alternative, and $r_{jk}^i=0$ and $r_{kj}^i=1$ if the $i$th decision-maker prefers the $k$th alternative to the $j$th alternative. This assumption implies that $r_{jk}^i+r_{kj}^i=1$, $\forall i,j,k$ and $a_i^j=\sum_{k=1}^m r_{jk}^i$ provides a valuation $R_j^i$ for the $j$th alternative. Then, Borda’s count is used to define feasible set $\textbf{F}$ as follows:
    \begin{equation}
        \textbf{F} = \Big\{ R_j^S | R_j^S \in \mathbb{R}^+, \hspace{2mm} 1 \leq R_j^S \leq m, \hspace{2mm}  \sum_{j=1}^m R_{j}^S = \frac{(m+1)m}{2} \Big\}
        \label{eq:setF}
    \end{equation}
    where, by convention, the number $m$ is assigned to the best alternative and number 1 to the worst. This definition is relevant as it is the key reason that \textbf{F} represents a convex set, hence ensuring feasibility in linear programming in direct contrast to other common social choice measures on rankings such as Kendall tau, Hamming and Kemeny distance functions.
    \item Information about preferences is ordinal and partial. In this case, some alternatives may be incomparable and this assumption implies that $r_{jk}^i+r_{kj}^i \leq 1$ $\forall i,j,k$. Now, $a_i^j$, defined as in the first case, establishes a minimum bound for $R_j^i$, and $b_i^j=m- \sum_{k=1}^m r_{jk}^i$ establishes a maximum bound for $R_j^i$. Then $R_j^i \in \left[a_i^j,b_i^j \right]$ and feasible set \textbf{F} is defined as in \eqref{eq:setF}.
    \item Information about preferences is cardinal and complete. In this particular case, a valued binary preference relation reports some degree of the cardinality of preferences from one alternative over another. Decision-makers' valuations should belong to a certain closed interval $\left[t_1,t_2 \right]$ as it was proposed by \cite{saaty1977scaling}, where it is also assumed that $R_{jk}^i=1/R_{kj}^i$. These assumptions lead to the following set \textbf{F}:
    \begin{equation}
        \textbf{F} = \Big\{ R_{jk}^S |  \hspace{2mm} t_1 \leq  R_{jk}^S \leq t_2 \Big\}.
    \end{equation}
\end{enumerate}

From the observation of \eqref{eq:Mink} and \eqref{eq:convex1}, a straightforward research question arises. If three different ethical principles can be considered by optimising a convex combination of different $p$-metric distance functions subject to some set of constraints, one may be able to explore a larger space of multiple ethical principles in a diverse society if additional $p$-metric distances are considered. In what follows, we further elaborate on this research question.

\section{A multi-$\ell_p$-norm approximation approach}
\label{sec:multinorm}
In this section, we propose a novel generalisation to the previous approach by  \cite{gonzalez1999distance,gonzalez2008aggregation,gonzalez2011design,gonzalez2016bentham} based on an \emph{$\ell_p$-norm approximation problem}, which is able to account for multiple, any-valued ethical principles.

Along the lines of \eqref{eq:convex1}, we now provide a formal definition of the multi-principle problem.
Our definition is conceived to provide expressiveness for: (i) each stakeholder to choose their own ethical principle; and (ii) setting the relative importance of ethical principles, and hence the preferences over ethical principles (e.g., three different groups supporting three ethical principles setting the relative importance by means of the size of the group).

\begin{definition}
We consider a non-negative weighted sum of $p$-metric distances, each representing the corresponding $U_p$ objective function.
Specifically, we consider a set $\textbf{P}$ of multiple ethical principles and a set $\boldsymbol{\lambda} \subseteq \mathbb{R}_{\geq 0}$, where $\lambda_p\in \boldsymbol{\lambda}$ represents the non-negative weight associated with ethical principle $p$.

We define the problem of computing the consensus $R^S$ considering the above-defined sum of ethical principles as the problem of computing
\begin{equation}\label{eq:multinorm}
    R^S=\argmin_{R} \sum_{p\in\textbf P}\lambda_p\left[  \sum_{i=1}^n \sum_{j=1}^m \sum_{k=1, k \neq j}^m w_i^p |R_{jk}^i-R_{jk} |^p\right]^{1/p}
\end{equation}
subject to the set of restrictions in $\textbf{F}$.
\end{definition}

We now show that solving \eqref{eq:multinorm} is equivalent to solving a \emph{multi-$\ell_p$-norm approximation problem}.
More specifically, we begin by defining the \emph{vectorisation} operation, which we will employ in Lemma \ref{le:single} to show that solving \eqref{eq:multinorm} for a single ethical principle is equivalent to solving a standard $\ell_p$-norm approximation problem \citep{boyd2004convex}.
Then, in Theorem \ref{th:multi} we leverage Lemma \ref{le:single} to show that solving \eqref{eq:multinorm} is equivalent to solving a \emph{multi-$\ell_p$-norm approximation problem}.

\begin{definition}
Given an $n\times m$ matrix $M$, the \emph{vectorisation} operation $\mathrm{vec}(M)$ produces a vector $v$ of $n\cdot m$ elements obtained by arranging the elements of $M$ in \emph{row-major} order, i.e., by arranging them sequentially row by row. Formally, $v_{i\cdot m+j}=M_{ij}$ $\forall i\in\{1,\ldots,n\}$ $\forall j\in\{1,\ldots,m\}$.
\end{definition}

\begin{lemma}\label{le:single}
The problem of computing the consensus
\begin{equation}\label{eq:singlenorm}
R^S=\argmin_{R} \left[  \sum_{i=1}^n \sum_{j=1}^m \sum_{k=1, k \neq j}^m w_i^p |R_{jk}^i-R_{jk} |^p\right]^{1/p}
\end{equation}
subject to $\textbf{F}$ is equivalent to solving the $\ell_p$-norm approximation problem
\begin{align}
\begin{split}\label{eq:nap}
    \text{minimise} \enspace &\overbrace{\lVert A x-b \rVert_p}^{\eta_p},\\
    \text{subject to} \enspace &\textbf F,
\end{split}
\end{align}
where
$$
A = \begin{bmatrix}
w_1\cdot I_{m^2} \\
\vdots \\
w_n\cdot I_{m^2}
\end{bmatrix}
,\quad
b = \begin{bmatrix}
w_1\cdot\mathrm{vec}(R^1) \\
\vdots \\
w_n\cdot\mathrm{vec}(R^n)
\end{bmatrix},
$$
and $I_{m^2}\in\mathbb{R}^{m^2\times m^2}$ is the identity matrix of size $m^2$.
For convenience, we denote as $\eta_p$ the value of the objective function of \eqref{eq:nap}, which will be employed later in \eqref{eq:weight}.
\end{lemma}
\begin{proof}
Without loss of generality, we assume that $R_{jk}=0$ when $j=k$. As a consequence, \eqref{eq:singlenorm} is equivalent to
\begin{equation}
    \argmin_{R^S} \left[  \sum_{i=1}^n \sum_{j=1}^{m^2} |w_i\cdot\mathrm{vec}(R^S)_j-w_i\cdot\mathrm{vec}(R^i)_j |^p\right]^{1/p}.
    \label{eq:neg_obj_1}
\end{equation}

We can further rewrite (\ref{eq:neg_obj_1}) as
\begin{equation}
    \argmin_{R^S}  \left[  \sum_{i=1}^n \lVert w_i\cdot\mathrm{vec}(R^S) -  w_i\cdot\mathrm{vec}(R^i) \rVert^p\right]^{1/p}.
    \label{eq:neg_obj_2}
\end{equation}

We define $A\in\mathbb{R}^{n \cdot m^2\times m^2}$ and $b\in\mathbb{R}^{n \cdot m^2}$ as
$$
A = \begin{bmatrix}
w_1\cdot I_{m^2} \\
\vdots \\
w_n\cdot I_{m^2}
\end{bmatrix}
,\quad
b = \begin{bmatrix}
w_1\cdot\mathrm{vec}(R^1) \\
\vdots \\
w_n\cdot\mathrm{vec}(R^n)
\end{bmatrix},
$$
where $I_{m^2}\in\mathbb{R}^{m^2\times m^2}$ is the identity matrix of size $m^2$.

By exploiting the above-defined $A$ and $b$, we can now formulate \eqref{eq:neg_obj_2} as an $\ell_p$-norm approximation problem, i.e.,
\begin{align}
    \text{minimise} \enspace &\lVert A x-b \rVert_p,\nonumber\\
     \text{subject to} \enspace &\textbf F.\nonumber
\end{align}
Notice that the solution to the $\ell_p$-norm approximation problem, i.e., the vector $x$, is the vectorisation of $R^S$ subject to $\textbf F$.
\end{proof}

Now we are ready to prove that computing the consensus for a set of ethical principals, as defined by \eqref{eq:multinorm}, amounts to solving a multi-$\ell_p$-norm approximation problem.

\begin{theorem}\label{th:multi}
We consider a set $\textbf{P}$ of multiple of ethical principles and a set $\boldsymbol{\lambda} \subseteq \mathbb{R}_{\geq 0}$, where $\lambda_p\in \boldsymbol{\lambda}$ represents the non-negative weight associated with ethical principle $p$.
The problem of computing the consensus
$$
R^S=\argmin_{R} \sum_{p\in\textbf P}\lambda_p\left[  \sum_{i=1}^n \sum_{j=1}^m \sum_{k=1, k \neq j}^m w_i^p |R_{jk}^i-R_{jk} |^p\right]^{1/p}
$$
subject to the set of restrictions in $\textbf{F}$ is equivalent to solving the multi-$\ell_p$-norm approximation problem
\begin{align}
\begin{split}\label{eq:pnap}
    \text{minimise} \enspace & \sum_{p\in\textbf P}\lambda_p \lVert A x-b \rVert_p,\\
    \text{subject to} \enspace &\textbf F,
\end{split}
\end{align}
where
$$
A = \begin{bmatrix}
w_1\cdot I_{m^2} \\
\vdots \\
w_n\cdot I_{m^2}
\end{bmatrix}
,\quad
b = \begin{bmatrix}
w_1\cdot\mathrm{vec}(R^1) \\
\vdots \\
w_n\cdot\mathrm{vec}(R^n)
\end{bmatrix}.
$$
and $I_{m^2}\in\mathbb{R}^{m^2\times m^2}$ is the identity matrix of size $m^2$.
\end{theorem}
\begin{proof}
By direct application of Lemma \ref{le:single} to each element of the sum.
\end{proof}

Notice that, while the concept of \emph{$\ell_p$-norm approximation problem} has been deeply studied in the optimisation literature \citep{boyd2004convex}, its generalisation to the \emph{multi-$\ell_p$-norm} case seems to have received very little attention.
To the best of our knowledge, the only work attempting to formalise and solve an $\ell_p$-norm approximation problem involving multiple norms is the one by \cite{samejima2016fast}, even though the approach proposed by the authors is not suitable to solve \eqref{eq:multinorm}.
Firstly, the \emph{general norm approximation problem} proposed in \cite{samejima2016fast} is not exactly equivalent to \eqref{eq:multinorm}, since it involves the minimisation of a sum of \emph{power of norms}, rather than considering norms directly. This is a formally different problem that is not aligned with the one proposed here, nor with previous literature \citep{gonzalez1999distance,gonzalez2008aggregation,gonzalez2011design,gonzalez2016bentham}.
Secondly, the algorithm proposed in \cite{samejima2016fast} is based on a numerical technique called \emph{Iteratively Reweighted Least Squares} (IRLS), which is not guaranteed to converge for $p\geq 3$.
Finally, such an algorithm can only be used to solve \emph{unconstrained} problems, i.e., it would disregard the restrictions in the set $\textbf{F}$.
These limitations further motivate the need of a general and reliable method to solve the consensus computation problem involving multiple ethical principles, such as the one presented here.

Thanks to Theorem \ref{th:multi} we can solve \eqref{eq:multinorm} by formulating it as in \eqref{eq:pnap}, hence enabling the use of Convex Optimisation solution techniques.
\begin{remark}\label{rem:convex}
If the restrictions in the set $\textbf F$ are convex, the problem in \eqref{eq:nap} is also convex, since it is well known that norms are convex functions \cite[Section 3.1.5]{boyd2004convex}.
Similarly, the problem in \eqref{eq:pnap} is also convex, since it is a non-negative weighted sum of convex functions \cite[Section 3.2.1]{boyd2004convex}.
\end{remark}
Specifically, we solve \eqref{eq:pnap} by employing the techniques discussed by \cite{Alizadeh}, Section 2.3g, who show how to represent inequalities involving $p$-norms as \emph{Second-Order Cone Programming} constraints, enabling the use of off-the-shelf solvers such as CPLEX.

\begin{remark}\label{rem:F}
The set of restrictions $\textbf F$ in \eqref{eq:multinorm} (or, equivalently, in \eqref{eq:pnap}) can be used to impose certain properties on the resulting consensus $R^S$, for instance to obtain a complete total ranking \citep{gonzalez2008aggregation} among the $m$ available options.
Imposing such restrictions inherently reduces the set of valid solutions, which may result in a consensus $R^S$ corresponding to a lower utility (as defined in \eqref{eq:Mink}) compared to the unrestricted case.
\end{remark}

\section{A re-weighting approach for multi-$\ell_p$-norm approximation}
\label{sec:weights}

The main purpose of \eqref{eq:pnap}, as well as of the original \eqref{eq:convex1} by \cite{gonzalez2016bentham}, is to compute a consensus $R^S$ that is the result of the aggregation of multiple ethical principles, or multiple $p$-norms.
Unfortunately, this approach alone does not always yield a result with the above-mentioned semantic, as illustrated by the following example.

Let us consider the example in which one aims at computing the multi-norm consensus for $\lambda_1=\lambda_2=0.5$ in \eqref{eq:convex1} or, equivalently, $\textbf{P}=\{1,\infty\}$ each weighted with the same $\lambda_p$ in \eqref{eq:pnap}.
Intuitively, the expected result should be a balanced mix of the single components (i.e., $p=1$ and $p=\infty$), since both are given the same importance (i.e., the same $\lambda$).

Nonetheless, directly summing $U_B$ and $U_R$ in \eqref{eq:convex1} disregards the fact that $U_B$ has a much higher value with respect to  $U_R$ (i.e., $184.17$ \emph{vs} $4$ in the example considered in our experiments) and hence, it receives a much higher importance in the optimisation.
Indeed, in this example the multi-norm consensus computed for $\mathbf{P}=\{1,\infty\}$ is \emph{exactly equal} to the one computed for the single-norm case with $p=1$.
In other words, the inclusion of $p=\infty$ has not a measurable impact in the resulting consensus, which is, of course, in contrast with the expected semantic.
More in general, this issue is due to the well known fact that different $p$-norms (or, equivalently, different $U_p$ functions) have different scales depending on $p$, which has no effect for the single-norm case, but must be taken into account for the multi-norm case.

To tackle this issue, we propose a \emph{re-weighting approach} for multi-$\ell_p$-norm approximation that aims at \emph{normalising} the values of the different $p$-norms. Let us first recall our definition of multi-$\ell_p$-norm approximation in \eqref{eq:pnap}:
\begin{align*}
\begin{split}
    \text{minimise} \enspace & \sum_{p\in\textbf P}\lambda_p \lVert A x-b \rVert_p,\\
    \text{subject to} \enspace &\textbf F.
\end{split}
\end{align*}
Our re-weighting approach seeks to determine the weight $\lambda_p\in\boldsymbol{\lambda}$ for each ethical principle $p\in\textbf P$ with the goal of minimising $\Psi(\textbf P,\boldsymbol{\lambda})$, i.e., the \emph{variance} of the set of values corresponding to each component $\lambda_p \lVert A x-b \rVert_p$, as defined hereafter.
\begin{definition}\label{def:psi}
Given a set of ethical principles $\textbf P$ and vector of weights $\boldsymbol{\lambda}$, we define the measure
\begin{equation}\label{eq:varp}
\Psi(\textbf P,\boldsymbol{\lambda})=Var(\{\lambda_p\lVert A x-b \rVert_p \mid p\in\textbf P,\lambda_p\in\boldsymbol{\lambda}\}),
\end{equation}
which quantifies how \emph{unbalanced} is the sum of the components corresponding to each ethical principle $p\in\textbf P$.
\end{definition}
With the goal of minimising the above-defined $\Psi$ measure, we propose the following re-weighting approach.

\begin{definition}
We define the \emph{re-weighted multi-$\ell_p$-norm approximation} problem as
\begin{align}
\begin{split}\label{eq:weight}
    \text{minimise} \enspace & \sum_{p\in\textbf P}\frac{1}{\eta_p}\cdot \lVert A x-b \rVert_p,\\
    \text{subject to} \enspace &\textbf F,
\end{split}
\end{align}
where $\eta_p$ is the value of the objective function of the single $\ell_p$-norm approximation problem in \eqref{eq:nap} considering  $A$ and $b$ (see previous definition in Lemma \ref{le:single}).
\end{definition}
In \eqref{eq:weight} each term of the multi-norm sum is divided for the value of $\lVert\cdot\rVert_p$ in the single-norm case, so as to normalise its contribution in the sum and, as a consequence, minimise the measure $\Psi$ that we defined as our objective.
Notice that computing each $\eta_p$ requires solving each single-norm approximation problem in \eqref{eq:nap} as a preliminary step to then solve \eqref{eq:weight}.

\begin{remark}
Following Remark \ref{rem:convex}, it is easy to show that \eqref{eq:weight} is also a convex optimisation problem if the restrictions in the set $\textbf F$ are convex.
\end{remark}

\section{Experimental results\label{sec:exp}}

In this section, we experimentally evaluate our approach following the methodology previously proposed in the literature \citep{gonzalez2016bentham}.
Specifically, we consider the participatory forest plan problem \citep{nordstrom2009aggregation} carried out in Lycksele (Sweden), where several stakeholders were questioned about their preferences with respect to several criteria.
Seven recreationists were interviewed about the relative importance attached to five criteria, following a pairwise comparison format. 
The original $5\times 5$ pairwise comparison matrices can be found in \cite{gonzalez2016bentham}.

Given the above experimental scenario, our objectives are the following:
\begin{enumerate}
    \item Analyse the distribution of the differences of the individual preferences with respect to the consensus (i.e., the residuals of the $\ell_p$-norm approximation problem) computed by using our model in \eqref{eq:nap} for several values of $p$.
    \item Analyse how considering different combinations of multiple ethical principles in our multi-$\ell_p$-norm approximation model impacts such a distribution of differences, both for the unweighted model (\eqref{eq:pnap}) and for the weighted one (\eqref{eq:weight}).\footnote{Our code is available at \url{https://github.com/filippobistaffa/social-choice-pnorm}.}
\end{enumerate}

\pgfplotsset{
    width=\textwidth,
    height=72mm,
}

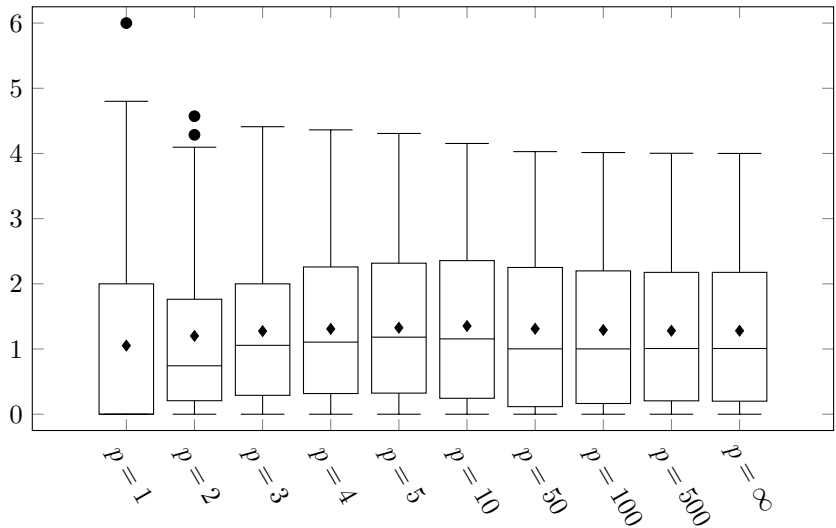
\begin{figure}[t]
    \centering
    \begin{tikzpicture}
        \begin{axis}[
            boxplot/draw direction=y,
            xtick={1,...,10},
            xticklabels={
                $p=1$,
                $p=2$,
                $p=3$,
                $p=4$,
                $p=5$,
                $p=10$,
                $p=50$,
                $p=100$,
                $p=500$,
                $p=\infty$
            },
            xticklabel style={
                rotate=-60,
                anchor=north west,
            },
            ymin=-0.25, ymax=6.25,
            ytick={0,...,6}
        ]
            \addplot [mark=*, boxplot={average=auto}]
            table [row sep=\\,y index=0] {
                data\\
                0.0000\\ 0.0600\\ 0.8600\\ 0.1900\\ 0.0000\\ 2.0000\\ 0.0000\\
                0.0000\\ 0.8000\\ 0.6700\\ 6.0000\\ 0.0000\\ 0.0000\\ 0.0000\\
                0.1900\\ 4.0000\\ 4.0000\\ 0.0000\\ 0.0000\\ 0.1900\\ 0.0000\\
                2.0000\\ 4.0000\\ 4.0000\\ 0.0000\\ 0.0000\\ 0.8000\\ 0.0000\\
                0.0000\\ 0.8000\\ 4.0000\\ 0.0000\\ 0.1300\\ 0.1300\\ 0.0000\\
                0.0000\\ 2.0000\\ 0.0000\\ 0.6700\\ 0.0000\\ 0.0000\\ 2.0000\\
                2.0000\\ 0.0000\\ 0.0000\\ 4.0000\\ 0.0000\\ 0.0000\\ 0.0000\\
                0.0000\\ 0.0000\\ 0.0000\\ 0.0000\\ 0.1900\\ 2.8000\\ 0.0000\\
                0.0000\\ 0.0000\\ 0.0600\\ 4.6700\\ 0.0000\\ 0.0000\\ 0.0000\\
                4.0000\\ 2.6700\\ 4.0000\\ 2.0000\\ 0.8000\\ 0.0000\\ 0.6700\\
                4.6700\\ 2.8000\\ 2.6700\\ 2.0000\\ 0.0000\\ 0.0000\\ 0.0900\\
                0.0000\\ 0.6700\\ 0.0000\\ 4.0000\\ 0.0000\\ 0.0000\\ 0.8000\\
                0.1300\\ 0.0000\\ 0.0000\\ 0.0000\\ 0.0000\\ 0.1900\\ 2.0000\\
                4.0000\\ 0.0000\\ 0.0000\\ 0.6700\\ 0.0000\\ 2.0000\\ 4.0000\\
                2.0000\\ 0.0000\\ 0.0000\\ 0.8000\\ 0.0000\\ 0.6700\\ 4.8000\\
                4.0000\\ 0.0000\\ 4.8000\\ 0.0000\\ 4.6700\\ 0.0000\\ 4.8000\\
                0.0000\\ 0.8000\\ 4.6700\\ 2.0000\\ 0.0000\\ 4.0000\\ 0.0000\\
                0.0000\\ 4.8000\\ 2.8000\\ 2.8000\\ 0.0000\\ 0.0000\\ 0.0000\\
                0.0000\\ 0.0000\\ 0.0000\\ 0.0000\\ 0.0000\\ 0.0000\\ 0.1300\\
                0.0000\\ 0.0000\\ 0.0000\\ 2.0000\\ 0.0000\\ 2.0000\\ 2.6700\\
                0.0000\\ 0.0000\\ 0.6700\\ 0.0000\\ 2.6700\\ 0.0000\\ 0.0000\\
                2.6700\\ 2.6700\\ 0.0000\\ 0.0000\\ 0.0000\\ 0.8000\\ 0.1300\\
                0.0600\\ 0.0000\\ 0.0000\\ 0.0000\\ 0.0000\\ 0.1300\\ 4.0000\\
                0.0000\\ 0.0000\\ 0.8000\\ 0.1900\\ 2.0000\\ 0.0000\\ 4.0000\\
                0.0000\\ 0.1300\\ 2.0000\\ 2.0000\\ 4.0000\\ 2.0000\\ 0.0000\\
            };
            \addplot [mark=*, boxplot={average=auto}]
            table [row sep=\\,y index=0] {
                data\\
                0.0000\\ 0.2671\\ 0.6228\\ 0.3085\\ 1.1914\\ 2.2859\\ 0.0000\\
                0.7228\\ 0.5615\\ 0.7228\\ 4.5715\\ 1.2573\\ 0.0000\\ 0.5328\\
                1.5385\\ 3.1430\\ 2.8570\\ 1.2185\\ 0.0000\\ 0.7171\\ 1.6387\\
                1.4856\\ 3.4487\\ 4.0958\\ 0.0000\\ 0.0000\\ 0.5929\\ 0.2372\\
                0.1185\\ 0.3914\\ 3.7141\\ 0.0000\\ 0.5928\\ 0.1085\\ 1.3928\\
                1.4285\\ 0.7427\\ 0.0000\\ 1.2028\\ 1.3485\\ 0.8570\\ 0.8570\\
                0.7815\\ 0.0000\\ 0.5271\\ 2.3613\\ 0.5144\\ 0.5513\\ 0.0958\\
                0.0000\\ 0.0000\\ 0.2071\\ 0.2372\\ 0.3085\\ 1.6086\\ 0.2859\\
                0.0000\\ 0.7228\\ 0.2985\\ 3.2772\\ 1.4285\\ 1.2573\\ 0.0000\\
                3.4672\\ 1.3215\\ 3.1430\\ 3.1430\\ 2.0185\\ 0.0000\\ 0.1429\\
                3.0313\\ 2.2856\\ 3.2213\\ 1.9042\\ 0.0000\\ 0.0000\\ 0.2971\\
                0.2372\\ 0.5515\\ 1.1914\\ 4.2859\\ 0.0000\\ 0.7228\\ 0.5615\\
                1.5228\\ 1.4285\\ 1.2573\\ 0.0000\\ 0.5328\\ 1.5385\\ 2.8570\\
                2.8570\\ 1.2185\\ 0.0000\\ 0.1429\\ 1.6387\\ 2.5144\\ 3.4487\\
                1.9042\\ 0.0000\\ 0.0000\\ 0.5929\\ 0.2372\\ 0.5515\\ 3.6086\\
                3.7141\\ 0.0000\\ 4.0772\\ 0.2385\\ 3.2772\\ 1.4285\\ 3.5427\\
                0.0000\\ 1.3328\\ 3.3215\\ 2.8570\\ 1.1430\\ 2.7815\\ 0.0000\\
                0.5271\\ 3.1613\\ 2.2856\\ 3.3513\\ 0.0958\\ 0.0000\\ 0.0000\\
                0.2071\\ 0.2372\\ 0.1185\\ 1.1914\\ 0.2859\\ 0.0000\\ 0.5928\\
                0.2385\\ 1.3928\\ 1.4285\\ 0.7427\\ 0.0000\\ 1.4672\\ 1.3215\\
                0.8570\\ 1.1430\\ 1.8885\\ 0.0000\\ 2.1429\\ 1.6387\\ 0.5144\\
                3.2213\\ 2.5742\\ 0.0000\\ 0.0000\\ 0.2071\\ 0.5628\\ 0.2485\\
                1.2514\\ 0.2859\\ 0.0000\\ 0.7228\\ 0.2385\\ 1.5228\\ 2.5715\\
                1.2573\\ 0.0000\\ 1.3328\\ 1.5385\\ 1.1430\\ 1.1430\\ 2.7815\\
                0.0000\\ 0.6571\\ 3.6387\\ 2.5144\\ 3.4487\\ 2.0958\\ 0.0000\\
            };
            \addplot [mark=*, boxplot={average=auto}]
            table [row sep=\\,y index=0] {
                data\\
                0.0000\\ 0.3500\\ 0.5383\\ 0.3793\\ 1.6259\\ 2.4104\\ 0.0000\\
                1.4213\\ 0.4856\\ 1.1740\\ 4.0000\\ 1.6154\\ 0.0000\\ 0.9433\\
                1.7712\\ 3.0334\\ 2.7335\\ 1.4324\\ 0.0000\\ 1.0195\\ 1.7396\\
                1.5070\\ 3.3669\\ 3.8120\\ 0.0000\\ 0.0000\\ 0.5100\\ 0.3217\\
                0.1893\\ 0.8259\\ 3.5896\\ 0.0000\\ 1.2913\\ 0.1844\\ 1.8440\\
                2.0000\\ 0.3846\\ 0.0000\\ 1.6133\\ 1.5812\\ 0.9666\\ 0.7335\\
                0.5676\\ 0.0000\\ 0.8295\\ 2.2604\\ 0.4930\\ 0.6331\\ 0.1880\\
                0.0000\\ 0.0000\\ 0.2900\\ 0.3217\\ 0.3793\\ 1.1741\\ 0.4104\\
                0.0000\\ 1.4213\\ 0.3744\\ 2.8260\\ 2.0000\\ 1.6154\\ 0.0000\\
                3.0567\\ 1.0888\\ 3.0334\\ 3.2665\\ 2.2324\\ 0.0000\\ 0.1595\\
                2.9304\\ 2.3070\\ 3.3031\\ 2.1880\\ 0.0000\\ 0.0000\\ 0.3800\\
                0.3217\\ 0.4807\\ 1.6259\\ 4.4104\\ 0.0000\\ 1.4213\\ 0.4856\\
                1.9740\\ 2.0000\\ 1.6154\\ 0.0000\\ 0.9433\\ 1.7712\\ 2.9666\\
                2.7335\\ 1.4324\\ 0.0000\\ 0.1595\\ 1.7396\\ 2.4930\\ 3.3669\\
                2.1880\\ 0.0000\\ 0.0000\\ 0.5100\\ 0.3217\\ 0.4807\\ 3.1741\\
                3.5896\\ 0.0000\\ 3.3787\\ 0.3144\\ 2.8260\\ 2.0000\\ 3.1846\\
                0.0000\\ 1.7433\\ 3.0888\\ 2.9666\\ 1.2665\\ 2.5676\\ 0.0000\\
                0.8295\\ 3.0604\\ 2.3070\\ 3.4331\\ 0.1880\\ 0.0000\\ 0.0000\\
                0.2900\\ 0.3217\\ 0.1893\\ 1.6259\\ 0.4104\\ 0.0000\\ 1.2913\\
                0.3144\\ 1.8440\\ 2.0000\\ 0.3846\\ 0.0000\\ 1.0567\\ 1.0888\\
                0.9666\\ 1.2665\\ 2.1024\\ 0.0000\\ 1.8405\\ 1.7396\\ 0.4930\\
                3.3031\\ 2.8580\\ 0.0000\\ 0.0000\\ 0.2900\\ 0.4783\\ 0.3193\\
                1.6859\\ 0.4104\\ 0.0000\\ 1.4213\\ 0.3144\\ 1.9740\\ 2.0000\\
                1.6154\\ 0.0000\\ 1.7433\\ 1.7712\\ 1.0334\\ 1.2665\\ 2.5676\\
                0.0000\\ 0.9595\\ 3.7396\\ 2.4930\\ 3.3669\\ 1.8120\\ 0.0000\\
            };
            \addplot [mark=*, boxplot={average=auto}]
            table [row sep=\\,y index=0] {
                data\\
                0.0000\\ 0.3773\\ 0.5094\\ 0.3998\\ 1.8419\\ 2.3617\\ 0.0000\\
                1.7303\\ 0.4618\\ 1.3339\\ 3.7411\\ 1.8553\\ 0.0000\\ 1.1479\\
                1.9290\\ 3.0086\\ 2.7303\\ 1.5273\\ 0.0000\\ 1.1660\\ 1.7200\\
                1.5490\\ 3.3585\\ 3.6543\\ 0.0000\\ 0.0000\\ 0.4827\\ 0.3506\\
                0.2098\\ 1.0419\\ 3.6383\\ 0.0000\\ 1.6003\\ 0.2082\\ 2.0039\\
                2.2589\\ 0.1447\\ 0.0000\\ 1.8179\\ 1.7390\\ 0.9914\\ 0.7303\\
                0.4727\\ 0.0000\\ 0.9760\\ 2.2800\\ 0.4510\\ 0.6415\\ 0.3457\\
                0.0000\\ 0.0000\\ 0.3173\\ 0.3506\\ 0.3998\\ 0.9581\\ 0.3617\\
                0.0000\\ 1.7303\\ 0.3982\\ 2.6661\\ 2.2589\\ 1.8553\\ 0.0000\\
                2.8521\\ 0.9310\\ 3.0086\\ 3.2697\\ 2.3273\\ 0.0000\\ 0.3060\\
                2.9500\\ 2.3490\\ 3.3115\\ 2.3457\\ 0.0000\\ 0.0000\\ 0.4073\\
                0.3506\\ 0.4602\\ 1.8419\\ 4.3617\\ 0.0000\\ 1.7303\\ 0.4618\\
                2.1339\\ 2.2589\\ 1.8553\\ 0.0000\\ 1.1479\\ 1.9290\\ 2.9914\\
                2.7303\\ 1.5273\\ 0.0000\\ 0.3060\\ 1.7200\\ 2.4510\\ 3.3585\\
                2.3457\\ 0.0000\\ 0.0000\\ 0.4827\\ 0.3506\\ 0.4602\\ 2.9581\\
                3.6383\\ 0.0000\\ 3.0697\\ 0.3382\\ 2.6661\\ 2.2589\\ 2.9447\\
                0.0000\\ 1.9479\\ 2.9310\\ 2.9914\\ 1.2697\\ 2.4727\\ 0.0000\\
                0.9760\\ 3.0800\\ 2.3490\\ 3.4415\\ 0.3457\\ 0.0000\\ 0.0000\\
                0.3173\\ 0.3506\\ 0.2098\\ 1.8419\\ 0.3617\\ 0.0000\\ 1.6003\\
                0.3382\\ 2.0039\\ 2.2589\\ 0.1447\\ 0.0000\\ 0.8521\\ 0.9310\\
                0.9914\\ 1.2697\\ 2.1973\\ 0.0000\\ 1.6940\\ 1.7200\\ 0.4510\\
                3.3115\\ 3.0157\\ 0.0000\\ 0.0000\\ 0.3173\\ 0.4494\\ 0.3398\\
                1.9019\\ 0.3617\\ 0.0000\\ 1.7303\\ 0.3382\\ 2.1339\\ 1.7411\\
                1.8553\\ 0.0000\\ 1.9479\\ 1.9290\\ 1.0086\\ 1.2697\\ 2.4727\\
                0.0000\\ 1.1060\\ 3.7200\\ 2.4510\\ 3.3585\\ 1.6543\\ 0.0000\\
            };
            \addplot [mark=*, boxplot={average=auto}]
            table [row sep=\\,y index=0] {
                data\\
                0.0000\\ 0.3837\\ 0.5020\\ 0.4032\\ 1.9725\\ 2.3066\\ 0.0000\\
                1.8958\\ 0.4575\\ 1.4129\\ 3.5821\\ 1.9882\\ 0.0000\\ 1.2687\\
                2.0422\\ 3.0023\\ 2.7628\\ 1.5737\\ 0.0000\\ 1.2420\\ 1.6732\\
                1.5727\\ 3.3577\\ 3.5554\\ 0.0000\\ 0.0000\\ 0.4763\\ 0.3580\\
                0.2132\\ 1.1725\\ 3.6934\\ 0.0000\\ 1.7658\\ 0.2125\\ 2.0829\\
                2.4179\\ 0.0118\\ 0.0000\\ 1.9387\\ 1.8522\\ 0.9977\\ 0.7628\\
                0.4263\\ 0.0000\\ 1.0520\\ 2.3268\\ 0.4273\\ 0.6423\\ 0.4446\\
                0.0000\\ 0.0000\\ 0.3237\\ 0.3580\\ 0.4032\\ 0.8275\\ 0.3066\\
                0.0000\\ 1.8958\\ 0.4025\\ 2.5871\\ 2.4179\\ 1.9882\\ 0.0000\\
                2.7313\\ 0.8178\\ 3.0023\\ 3.2372\\ 2.3737\\ 0.0000\\ 0.3820\\
                2.9968\\ 2.3727\\ 3.3123\\ 2.4446\\ 0.0000\\ 0.0000\\ 0.4137\\
                0.3580\\ 0.4568\\ 1.9725\\ 4.3066\\ 0.0000\\ 1.8958\\ 0.4575\\
                2.2129\\ 2.4179\\ 1.9882\\ 0.0000\\ 1.2687\\ 2.0422\\ 2.9977\\
                2.7628\\ 1.5737\\ 0.0000\\ 0.3820\\ 1.6732\\ 2.4273\\ 3.3577\\
                2.4446\\ 0.0000\\ 0.0000\\ 0.4763\\ 0.3580\\ 0.4568\\ 2.8275\\
                3.6934\\ 0.0000\\ 2.9042\\ 0.3425\\ 2.5871\\ 2.4179\\ 2.8118\\
                0.0000\\ 2.0687\\ 2.8178\\ 2.9977\\ 1.2372\\ 2.4263\\ 0.0000\\
                1.0520\\ 3.1268\\ 2.3727\\ 3.4423\\ 0.4446\\ 0.0000\\ 0.0000\\
                0.3237\\ 0.3580\\ 0.2132\\ 1.9725\\ 0.3066\\ 0.0000\\ 1.7658\\
                0.3425\\ 2.0829\\ 2.4179\\ 0.0118\\ 0.0000\\ 0.7313\\ 0.8178\\
                0.9977\\ 1.2372\\ 2.2437\\ 0.0000\\ 1.6180\\ 1.6732\\ 0.4273\\
                3.3123\\ 3.1146\\ 0.0000\\ 0.0000\\ 0.3237\\ 0.4420\\ 0.3432\\
                2.0325\\ 0.3066\\ 0.0000\\ 1.8958\\ 0.3425\\ 2.2129\\ 1.5821\\
                1.9882\\ 0.0000\\ 2.0687\\ 2.0422\\ 1.0023\\ 1.2372\\ 2.4263\\
                0.0000\\ 1.1820\\ 3.6732\\ 2.4273\\ 3.3577\\ 1.5554\\ 0.0000\\
            };
            \addplot [mark=*, boxplot={average=auto}]
            table [row sep=\\,y index=0] {
                data\\
                0.0000\\ 0.3045\\ 0.5805\\ 0.3366\\ 2.2061\\ 2.1533\\ 0.0000\\
                2.1819\\ 0.5304\\ 1.5365\\ 3.2675\\ 2.2154\\ 0.0000\\ 1.4710\\
                2.2628\\ 3.0000\\ 2.8846\\ 1.6244\\ 0.0000\\ 1.3454\\ 1.5083\\
                1.5982\\ 3.3590\\ 3.3775\\ 0.0000\\ 0.0000\\ 0.5555\\ 0.2795\\
                0.1466\\ 1.4061\\ 3.8467\\ 0.0000\\ 2.0519\\ 0.1396\\ 2.2065\\
                2.7325\\ 0.2154\\ 0.0000\\ 2.1410\\ 2.0728\\ 1.0000\\ 0.8846\\
                0.3756\\ 0.0000\\ 1.1554\\ 2.4917\\ 0.4018\\ 0.6410\\ 0.6225\\
                0.0000\\ 0.0000\\ 0.2445\\ 0.2795\\ 0.3366\\ 0.5939\\ 0.1533\\
                0.0000\\ 2.1819\\ 0.3296\\ 2.4635\\ 2.7325\\ 2.2154\\ 0.0000\\
                2.5290\\ 0.5972\\ 3.0000\\ 3.1154\\ 2.4244\\ 0.0000\\ 0.4854\\
                3.1617\\ 2.3982\\ 3.3110\\ 2.6225\\ 0.0000\\ 0.0000\\ 0.3345\\
                0.2795\\ 0.5234\\ 2.2061\\ 4.1533\\ 0.0000\\ 2.1819\\ 0.5304\\
                2.3365\\ 2.7325\\ 2.2154\\ 0.0000\\ 1.4710\\ 2.2628\\ 3.0000\\
                2.8846\\ 1.6244\\ 0.0000\\ 0.4854\\ 1.5083\\ 2.4018\\ 3.3590\\
                2.6225\\ 0.0000\\ 0.0000\\ 0.5555\\ 0.2795\\ 0.5234\\ 2.5939\\
                3.8467\\ 0.0000\\ 2.6181\\ 0.2696\\ 2.4635\\ 2.7325\\ 2.5846\\
                0.0000\\ 2.2710\\ 2.5972\\ 3.0000\\ 1.1154\\ 2.3756\\ 0.0000\\
                1.1554\\ 3.2917\\ 2.3982\\ 3.4410\\ 0.6225\\ 0.0000\\ 0.0000\\
                0.2445\\ 0.2795\\ 0.1466\\ 2.2061\\ 0.1533\\ 0.0000\\ 2.0519\\
                0.2696\\ 2.2065\\ 2.7325\\ 0.2154\\ 0.0000\\ 0.5290\\ 0.5972\\
                1.0000\\ 1.1154\\ 2.2944\\ 0.0000\\ 1.5146\\ 1.5083\\ 0.4018\\
                3.3110\\ 3.2925\\ 0.0000\\ 0.0000\\ 0.2445\\ 0.5205\\ 0.2766\\
                2.2661\\ 0.1533\\ 0.0000\\ 2.1819\\ 0.2696\\ 2.3365\\ 1.2675\\
                2.2154\\ 0.0000\\ 2.2710\\ 2.2628\\ 1.0000\\ 1.1154\\ 2.3756\\
                0.0000\\ 1.2854\\ 3.5083\\ 2.4018\\ 3.3590\\ 1.3775\\ 0.0000\\
            };
            \addplot [mark=*, boxplot={average=auto}]
            table [row sep=\\,y index=0] {
                data\\
                0.0000\\ 0.2677\\ 0.5995\\ 0.3061\\ 1.9602\\ 2.0283\\ 0.0000\\
                1.8910\\ 0.5630\\ 1.3621\\ 3.1659\\ 1.9822\\ 0.0000\\ 1.2117\\
                2.0574\\ 3.0025\\ 2.9304\\ 1.5161\\ 0.0000\\ 0.8956\\ 1.4059\\
                1.5566\\ 3.3711\\ 3.3361\\ 0.0000\\ 0.0000\\ 0.5923\\ 0.2605\\
                0.1161\\ 1.1602\\ 3.9717\\ 0.0000\\ 1.7610\\ 0.1070\\ 2.0321\\
                2.8341\\ 0.0178\\ 0.0000\\ 1.8817\\ 1.8674\\ 0.9975\\ 0.9304\\
                0.4839\\ 0.0000\\ 0.7056\\ 2.5941\\ 0.4434\\ 0.6289\\ 0.6639\\
                0.0000\\ 0.0000\\ 0.2077\\ 0.2605\\ 0.3061\\ 0.8398\\ 0.0283\\
                0.0000\\ 1.8910\\ 0.2970\\ 2.6379\\ 2.8341\\ 1.9822\\ 0.0000\\
                2.7883\\ 0.8026\\ 3.0025\\ 3.0696\\ 2.3161\\ 0.0000\\ 0.0356\\
                3.2641\\ 2.3566\\ 3.2989\\ 2.6639\\ 0.0000\\ 0.0000\\ 0.2977\\
                0.2605\\ 0.5539\\ 1.9602\\ 4.0283\\ 0.0000\\ 1.8910\\ 0.5630\\
                2.1621\\ 2.8341\\ 1.9822\\ 0.0000\\ 1.2117\\ 2.0574\\ 2.9975\\
                2.9304\\ 1.5161\\ 0.0000\\ 0.0356\\ 1.4059\\ 2.4434\\ 3.3711\\
                2.6639\\ 0.0000\\ 0.0000\\ 0.5923\\ 0.2605\\ 0.5539\\ 2.8398\\
                3.9717\\ 0.0000\\ 2.9090\\ 0.2370\\ 2.6379\\ 2.8341\\ 2.8178\\
                0.0000\\ 2.0117\\ 2.8026\\ 2.9975\\ 1.0696\\ 2.4839\\ 0.0000\\
                0.7056\\ 3.3941\\ 2.3566\\ 3.4289\\ 0.6639\\ 0.0000\\ 0.0000\\
                0.2077\\ 0.2605\\ 0.1161\\ 1.9602\\ 0.0283\\ 0.0000\\ 1.7610\\
                0.2370\\ 2.0321\\ 2.8341\\ 0.0178\\ 0.0000\\ 0.7883\\ 0.8026\\
                0.9975\\ 1.0696\\ 2.1861\\ 0.0000\\ 1.9644\\ 1.4059\\ 0.4434\\
                3.2989\\ 3.3339\\ 0.0000\\ 0.0000\\ 0.2077\\ 0.5395\\ 0.2461\\
                2.0202\\ 0.0283\\ 0.0000\\ 1.8910\\ 0.2370\\ 2.1621\\ 1.1659\\
                1.9822\\ 0.0000\\ 2.0117\\ 2.0574\\ 1.0025\\ 1.0696\\ 2.4839\\
                0.0000\\ 0.8356\\ 3.4059\\ 2.4434\\ 3.3711\\ 1.3361\\ 0.0000\\
            };
            \addplot [mark=*, boxplot={average=auto}]
            table [row sep=\\,y index=0] {
                data\\
                0.0000\\ 0.2666\\ 0.6003\\ 0.3053\\ 1.8020\\ 2.0140\\ 0.0000\\
                1.6892\\ 0.5639\\ 1.2623\\ 3.3596\\ 1.8356\\ 0.0000\\ 1.0693\\
                1.9298\\ 3.0169\\ 2.8589\\ 1.4636\\ 0.0000\\ 0.8645\\ 1.4425\\
                1.5353\\ 3.3662\\ 3.3895\\ 0.0000\\ 0.0000\\ 0.5934\\ 0.2597\\
                0.1153\\ 1.0020\\ 3.9860\\ 0.0000\\ 1.5592\\ 0.1061\\ 1.9323\\
                2.6404\\ 0.1644\\ 0.0000\\ 1.7393\\ 1.7398\\ 0.9831\\ 0.8589\\
                0.5364\\ 0.0000\\ 0.6745\\ 2.5575\\ 0.4647\\ 0.6338\\ 0.6105\\
                0.0000\\ 0.0000\\ 0.2066\\ 0.2597\\ 0.3053\\ 0.9980\\ 0.0140\\
                0.0000\\ 1.6892\\ 0.2961\\ 2.7377\\ 2.6404\\ 1.8356\\ 0.0000\\
                2.9307\\ 0.9302\\ 3.0169\\ 3.1411\\ 2.2636\\ 0.0000\\ 0.0045\\
                3.2275\\ 2.3353\\ 3.3038\\ 2.6105\\ 0.0000\\ 0.0000\\ 0.2966\\
                0.2597\\ 0.5547\\ 1.8020\\ 4.0140\\ 0.0000\\ 1.6892\\ 0.5639\\
                2.0623\\ 2.6404\\ 1.8356\\ 0.0000\\ 1.0693\\ 1.9298\\ 2.9831\\
                2.8589\\ 1.4636\\ 0.0000\\ 0.0045\\ 1.4425\\ 2.4647\\ 3.3662\\
                2.6105\\ 0.0000\\ 0.0000\\ 0.5934\\ 0.2597\\ 0.5547\\ 2.9980\\
                3.9860\\ 0.0000\\ 3.1108\\ 0.2361\\ 2.7377\\ 2.6404\\ 2.9644\\
                0.0000\\ 1.8693\\ 2.9302\\ 2.9831\\ 1.1411\\ 2.5364\\ 0.0000\\
                0.6745\\ 3.3575\\ 2.3353\\ 3.4338\\ 0.6105\\ 0.0000\\ 0.0000\\
                0.2066\\ 0.2597\\ 0.1153\\ 1.8020\\ 0.0140\\ 0.0000\\ 1.5592\\
                0.2361\\ 1.9323\\ 2.6404\\ 0.1644\\ 0.0000\\ 0.9307\\ 0.9302\\
                0.9831\\ 1.1411\\ 2.1336\\ 0.0000\\ 1.9955\\ 1.4425\\ 0.4647\\
                3.3038\\ 3.2805\\ 0.0000\\ 0.0000\\ 0.2066\\ 0.5403\\ 0.2453\\
                1.8620\\ 0.0140\\ 0.0000\\ 1.6892\\ 0.2361\\ 2.0623\\ 1.3596\\
                1.8356\\ 0.0000\\ 1.8693\\ 1.9298\\ 1.0169\\ 1.1411\\ 2.5364\\
                0.0000\\ 0.8045\\ 3.4425\\ 2.4647\\ 3.3662\\ 1.3895\\ 0.0000\\
            };
            \addplot [mark=*, boxplot={average=auto}]
            table [row sep=\\,y index=0] {
                data\\
                0.0000\\ 0.2653\\ 0.6020\\ 0.3044\\ 1.7322\\ 2.0028\\ 0.0000\\
                1.5886\\ 0.5650\\ 1.2180\\ 3.5111\\ 1.7741\\ 0.0000\\ 1.0078\\
                1.8797\\ 3.0284\\ 2.8288\\ 1.4408\\ 0.0000\\ 0.8423\\ 1.5219\\
                1.5273\\ 3.3657\\ 3.4819\\ 0.0000\\ 0.0000\\ 0.5947\\ 0.2580\\
                0.1144\\ 0.9322\\ 3.9972\\ 0.0000\\ 1.4586\\ 0.1050\\ 1.8880\\
                2.4889\\ 0.2259\\ 0.0000\\ 1.6778\\ 1.6897\\ 0.9716\\ 0.8288\\
                0.5592\\ 0.0000\\ 0.6523\\ 2.4781\\ 0.4727\\ 0.6343\\ 0.5181\\
                0.0000\\ 0.0000\\ 0.2053\\ 0.2580\\ 0.3044\\ 1.0678\\ 0.0028\\
                0.0000\\ 1.5886\\ 0.2950\\ 2.7820\\ 2.4889\\ 1.7741\\ 0.0000\\
                2.9922\\ 0.9803\\ 3.0284\\ 3.1712\\ 2.2408\\ 0.0000\\ 0.0177\\
                3.1481\\ 2.3273\\ 3.3043\\ 2.5181\\ 0.0000\\ 0.0000\\ 0.2953\\
                0.2580\\ 0.5556\\ 1.7322\\ 4.0028\\ 0.0000\\ 1.5886\\ 0.5650\\
                2.0180\\ 2.4889\\ 1.7741\\ 0.0000\\ 1.0078\\ 1.8797\\ 2.9716\\
                2.8288\\ 1.4408\\ 0.0000\\ 0.0177\\ 1.5219\\ 2.4727\\ 3.3657\\
                2.5181\\ 0.0000\\ 0.0000\\ 0.5947\\ 0.2580\\ 0.5556\\ 3.0678\\
                3.9972\\ 0.0000\\ 3.2114\\ 0.2350\\ 2.7820\\ 2.4889\\ 3.0259\\
                0.0000\\ 1.8078\\ 2.9803\\ 2.9716\\ 1.1712\\ 2.5592\\ 0.0000\\
                0.6523\\ 3.2781\\ 2.3273\\ 3.4343\\ 0.5181\\ 0.0000\\ 0.0000\\
                0.2053\\ 0.2580\\ 0.1144\\ 1.7322\\ 0.0028\\ 0.0000\\ 1.4586\\
                0.2350\\ 1.8880\\ 2.4889\\ 0.2259\\ 0.0000\\ 0.9922\\ 0.9803\\
                0.9716\\ 1.1712\\ 2.1108\\ 0.0000\\ 2.0177\\ 1.5219\\ 0.4727\\
                3.3043\\ 3.1881\\ 0.0000\\ 0.0000\\ 0.2053\\ 0.5420\\ 0.2444\\
                1.7922\\ 0.0028\\ 0.0000\\ 1.5886\\ 0.2350\\ 2.0180\\ 1.5111\\
                1.7741\\ 0.0000\\ 1.8078\\ 1.8797\\ 1.0284\\ 1.1712\\ 2.5592\\
                0.0000\\ 0.7823\\ 3.5219\\ 2.4727\\ 3.3657\\ 1.4819\\ 0.0000\\
            };
            \addplot [mark=*, boxplot={average=auto}]
            table [row sep=\\,y index=0] {
                data\\
                0.0000\\ 0.2600\\ 0.6020\\ 0.3040\\ 1.7320\\ 2.0000\\ 0.0000\\
                1.5890\\ 0.5650\\ 1.2180\\ 3.5110\\ 1.7740\\ 0.0000\\ 1.0080\\
                1.8800\\ 3.0280\\ 2.8290\\ 1.4410\\ 0.0000\\ 0.8420\\ 1.5220\\
                1.5270\\ 3.3660\\ 3.4820\\ 0.0000\\ 0.0000\\ 0.6000\\ 0.2580\\
                0.1140\\ 0.9320\\ 4.0000\\ 0.0000\\ 1.4590\\ 0.1050\\ 1.8880\\
                2.4890\\ 0.2260\\ 0.0000\\ 1.6780\\ 1.6900\\ 0.9720\\ 0.8290\\
                0.5590\\ 0.0000\\ 0.6520\\ 2.4780\\ 0.4730\\ 0.6340\\ 0.5180\\
                0.0000\\ 0.0000\\ 0.2000\\ 0.2580\\ 0.3040\\ 1.0680\\ 0.0000\\
                0.0000\\ 1.5890\\ 0.2950\\ 2.7820\\ 2.4890\\ 1.7740\\ 0.0000\\
                2.9920\\ 0.9800\\ 3.0280\\ 3.1710\\ 2.2410\\ 0.0000\\ 0.0180\\
                3.1480\\ 2.3270\\ 3.3040\\ 2.5180\\ 0.0000\\ 0.0000\\ 0.2900\\
                0.2580\\ 0.5560\\ 1.7320\\ 4.0000\\ 0.0000\\ 1.5890\\ 0.5650\\
                2.0180\\ 2.4890\\ 1.7740\\ 0.0000\\ 1.0080\\ 1.8800\\ 2.9720\\
                2.8290\\ 1.4410\\ 0.0000\\ 0.0180\\ 1.5220\\ 2.4730\\ 3.3660\\
                2.5180\\ 0.0000\\ 0.0000\\ 0.6000\\ 0.2580\\ 0.5560\\ 3.0680\\
                4.0000\\ 0.0000\\ 3.2110\\ 0.2350\\ 2.7820\\ 2.4890\\ 3.0260\\
                0.0000\\ 1.8080\\ 2.9800\\ 2.9720\\ 1.1710\\ 2.5590\\ 0.0000\\
                0.6520\\ 3.2780\\ 2.3270\\ 3.4340\\ 0.5180\\ 0.0000\\ 0.0000\\
                0.2000\\ 0.2580\\ 0.1140\\ 1.7320\\ 0.0000\\ 0.0000\\ 1.4590\\
                0.2350\\ 1.8880\\ 2.4890\\ 0.2260\\ 0.0000\\ 0.9920\\ 0.9800\\
                0.9720\\ 1.1710\\ 2.1110\\ 0.0000\\ 2.0180\\ 1.5220\\ 0.4730\\
                3.3040\\ 3.1880\\ 0.0000\\ 0.0000\\ 0.2000\\ 0.5420\\ 0.2440\\
                1.7920\\ 0.0000\\ 0.0000\\ 1.5890\\ 0.2350\\ 2.0180\\ 1.5110\\
                1.7740\\ 0.0000\\ 1.8080\\ 1.8800\\ 1.0280\\ 1.1710\\ 2.5590\\
                0.0000\\ 0.7820\\ 3.5220\\ 2.4730\\ 3.3660\\ 1.4820\\ 0.0000\\
            };
        \end{axis}
    \end{tikzpicture}
    \caption{Boxplots representing the distribution of residuals for different ethical principles.}
    \label{fig:single}
\end{figure}

\subsection{Single ethical principle case}

One of the main contributions of our work is generalising previous consensus computation approaches \citep{gonzalez2016bentham} by providing the opportunity of considering \emph{any} ethical principle $p$ (and not only $p\in\{1,\infty\}$).
Along these lines, in our first experiment we compute the consensus by using our model in \eqref{eq:nap} for $p\in\{1,2,3,4,5,10,50,100,500,\infty\}$.
Then, in Figure \ref{fig:single} we show the boxplots of the distribution of the differences of the individual preferences with respect to the consensus (i.e., the residuals of the corresponding $\ell_p$-norm approximation problem) computed for each $p$.
Here we report such boxplots (rather than reporting the actual consensus matrix $R^S$) because they allow us to compare the distributions of the residuals in a visual way for a large number of ethical principles, which is the goal of this experiment.

As expected, $p=1$ (i.e., fully utilitarian case) is the ethical principle that minimises the \emph{average} difference ($\mu_1 \simeq 1.05$) among all possible choices, since considering the $\ell_1$-norm amounts to minimising the \emph{total} sum of differences and, hence, the average one as well.
The $\ell_1$-norm is also associated with the highest \emph{maximum} difference ($\text{max}_1=6$).

On the other hand, $p=2$ results in the consensus that minimises the \emph{mean squared error}, i.e., it results in the distribution of differences with the lowest \emph{variance} ($\sigma^2_2=1.45$, see Table \ref{tab:multi}), corresponding to the boxplot with the lowest \emph{interquartile range} among all ethical principles.

Intuitively, we can interpret this result as follows:
for $p=2$ the individual differences are more compactly distributed, i.e., they are less sparsely distributed.
More in general, $p=2$ can be interpreted as the consensus that aims at the most \emph{homogeneous} difference across all individuals, rather than simply minimising the maximum one (as done for $p=\infty$).

Moreover, we immediately observe that increasing $p$ (i.e., shifting to a more egalitarian consensus) results in a \emph{lower} maximum difference ($\text{max}_2=4.57$) with respect to $p=1$.
Indeed, as we increase $p$, the maximum difference consistently decreases until reaching $p=\infty$ (i.e., fully egalitarian), whose objective indeed is minimising such a maximum difference.
Nonetheless, our results also show that the differences among the ethical principles for $p\geq 3$ are not very significant.


\subsection{Multiple ethical principle case}

In this second experiment we investigate how considering multiple ethical principles impacts the distribution of residuals, along the lines of the methodology followed in the previous section. Furthermore, in this case we are interested in evaluating whether our re-weighting approach achieves our goal of balancing the contributions of the different ethical principles in terms of the measure $\Psi$ in Definition \ref{def:psi}.
To this end, we consider all the possible subsets of $\{1,2,\infty\}$ as our set $\mathbf{P}$ of ethical principles.\footnote{We also consider subsets of size 1 (i.e., cases where $\mathbf{P}$ contains only a single ethical principle), which will serve as a reference to better evaluate the multi-norm cases.}
Notice that our approach would allow us to consider any arbitrary set $\mathbf{P}$ of ethical principles. 
Nonetheless, here we devote our attention to $p\in\{1,2,\infty\}$ as these ethical principles can be semantically characterised in a clear way, as showed by the results of our previous experiment.
Specifically, $p=1$ results in a consensus that minimises the \emph{average} residual, $p=2$ to one that minimises the \emph{standard deviation} of the residuals, and $p=\infty$ to one that minimises the \emph{maximum} residual.

In particular, we are interested in evaluating the effect of the reweighting scheme proposed in Section \ref{sec:weights} on the metrics that characterise the distribution of the residuals.
Hence, in Table \ref{tab:multi} we compare the descriptive statistics (namely the \emph{average} and the \emph{standard deviation}, as well as the \emph{minimum} and \emph{maximum}) of the distribution of the residuals obtained by solving \eqref{eq:pnap} (i.e., without reweighting) and \eqref{eq:weight} (i.e., with reweighting).

We immediately observe that the consensus computed by solving \eqref{eq:pnap} for all the cases that include $p=1$ (i.e., $\mathbf{P}=\{1,2\}$, $\mathbf{P}=\{1,\infty\}$, and $\mathbf{P}=\{1,2,\infty\}$) is exactly equal to the one computed for $p=1$ in the single-norm case, experimentally confirming our discussion in Section \ref{sec:weights}.
This is also quantitatively confirmed by the extremely high values of the measure $\Psi$, which indicates that the contributions of the different ethical principles are very unbalanced.
Based on these results, we conclude that \eqref{eq:pnap} is not suitable to compute a consensus based on a set $\mathbf{P}$ of multiple ethical principles including $p=1$, since such a $p$ dominates any other $p$.

On the other hand, we observe that the multi-norm consensus obtained with our reweighting scheme (\eqref{eq:weight}) is characterised by descriptive statistics that are a balanced mix of its single components.
In fact, as expected, the $\Psi$ measure (i.e., the variance) is very close to zero, indicating that the contributions of the different ethical principles are practically equal.
For instance, for $\mathbf{P}=\{1,2\}$ we observe a lower $\sigma^2$ compared to $p=1$ and a lower $\mu$ compared to $p=2$.
That is, we correctly observe the ``contributions'' of both $p=1$ (i.e., reducing $\mu$) and $p=2$ (i.e., reducing $\sigma^2$).
Similarly, for $\mathbf{P}=\{1,\infty\}$ we observe a lower maximum residual compared to $p=1$ and a lower $\mu$ compared to $p=\infty$, and for $\mathbf{P}=\{2,\infty\}$ we observe a lower maximum residual compared to $p=2$ and a lower $\sigma^2$ compared to $p=\infty$.
Finally, when we combine all three ethical principles, we correctly obtain a consensus where $\mu$ is lower compared to $p=2$ and $p=\infty$, $\sigma^2$ is lower compared to $p=1$ and $p=\infty$, and maximum residual is lower compared to $p=1$ and $p=2$.

\def\fcw{12mm}

\begin{table}[ht]
\setlength\tabcolsep{0.5pt}
\centering
\footnotesize
\begin{tabularx}{\columnwidth}{>{\centering\arraybackslash}p{\fcw} XXXXXXXXXX}
\toprule
& \multicolumn{5}{c}{\eqref{eq:pnap} (no reweighting)} & \multicolumn{5}{c}{\eqref{eq:weight} (reweighting)} \\
{$\mathbf{P}$} & {min} & {max} & {$\mu$} & {$\sigma^2$} & {$\Psi$} & {min} & {max} & {$\mu$} & {$\sigma^2$} & $\Psi$ \\
\midrule
$\{1\}$ &
\num{0.00000000} & \num{6.00000000} & \num{1.05240000} & \num{2.46956432} & \num{0.00000000} &
\num{0.00000000} & \num{6.00000000} & \num{1.05240000} & \num{2.46956432} & \num{0.00000000} \\
$\{2\}$ &
\num{0.00000000} & \num{4.57154291} & \num{1.20064981} & \num{1.45278444} & \num{0.00000000} &
\num{0.00000000} & \num{4.57154291} & \num{1.20064981} & \num{1.45278444} & \num{0.00000000} \\
$\{\infty\}$ &
\num{0.00000000} & \num{4.00000000} & \num{3.02680000} & \num{1.74891384} & \num{0.00000000} &
\num{0.00000000} & \num{4.00000000} & \num{3.02680000} & \num{1.74891384} & \num{0.00000000} \\
$\{1,2\}$ &
\num{0.00000000} & \num{5.99999999} & \num{1.05240000} & \num{2.46956429} & \num{6336.1} &
\num{0.00000000} & \num{5.78293340} & \num{1.09617716} & \num{1.87582216} & \num{0.00002901} \\
$\{1,\infty\}$ &
\num{0.00000000} & \num{6.00000000} & \num{1.05240000} & \num{2.46956432} & \num{7936.1} &
\num{0.00000000} & \num{4.00000000} & \num{1.12794286} & \num{1.91665551} & \num{0.00128815} \\
$\{2,\infty\}$ &
\num{0.00000000} & \num{4.00000003} & \num{1.20720811} & \num{1.45393060} & \num{85.92648846} &
\num{0.00000000} & \num{4.00000040} & \num{1.20720845} & \num{1.45393063} & \num{0.00000214} \\
$\{1,2,\infty\}$ &
\num{0.00000000} & \num{5.99999995} & \num{1.05240001} & \num{2.46956423} & \num{6383.2} &
\num{0.00000000} & \num{4.00000017} & \num{1.14615652} & \num{1.70176711} & \num{0.00145101} \\
\bottomrule
\end{tabularx}
\caption{\label{tab:multi}Descriptive statistics for the residulas with all $\mathbf{P}\subseteq\{1,2,\infty\}$. Notice that for single-norm cases (first 3 rows) \eqref{eq:pnap} and \eqref{eq:weight} correctly provide the same results, since reweighting has no impact when $\mathbf{P}$ contains only one $p$.}
\end{table}

\section{Discussion \& Concluding Remarks\label{sec:conc}}

In this paper, we formulated and solved the problem of computing a consensus based on multiple ethical principles among the preferences of different individuals so that the ethical principles involved correctly contribute to such a consensus. Furthermore, we empirically analysed the impact of different configurations of single and multiple ethical principles on the resulting consensus.  

The implications derived from this paper are twofold. From a theoretical perspective, using a $p$-metric as a proxy to represent a criterion or an ethical principle to reach a consensus allows us to extend previous analysis to account for multiple ethical principles that represent the actual diversity in individuals and social groups. From a practical perspective, we reformulate the problem to use a state-of-the-art algorithm that allows us to explore the characteristics of the possible consensus when multiple perspectives, not only the extreme values, are considered. 

We should highlight that the choice of parameter $p$ may be open to debate. Although $p$ can be mathematically interpreted as the positioning between the principle of maximum freedom and the principle of maximum fairness, values of $p$ different from 1, 2, and $\infty$ have a complex semantic interpretation.
This conclusion has been possible thanks to our novel technique to compute the consensus for any value of $p$, including the intermediate ones (which, based on our knowledge of the state of the art, was not possible before). 
Along these lines, even if we could not provide a precise semantic interpretation of each ethical principle $p$, our contributions allowed us to conclude that any $p \geq 3$ is not significantly different from $p=\infty$. We also remark that our conclusions are bound to the particular dataset that we considered in our experiments, which might vary in other test cases. On the contrary, parameters $\lambda$ are easier to interpret and assign. Since each value of $\lambda$ stands for the social influence, or importance, of each particular individual (or group of people) involved in the group decision-making, it is easier to conduct sensitivity analysis or explore solutions compared to~$p$.

Our work represents further motivation to enhance the integration of multiple ethical principles in group decision-making. 
First, it might be worth considering different distance functions other than $p$-metric distances (adopted both in this paper and in \cite{gonzalez2016bentham}), as well as studying the computational properties of the associated consensus computation problem. Second, our contribution allows us to compare our general approach to aggregate judgements with existing algorithms in the ranking aggregation literature to aggregate different rankings. This might spur interesting cross-fertilisation between these two different but related problems.

Finally, our work opens the door to \emph{collective ethics}. 
As argued in \cite{gabriel2020artificial}, we live in a pluralistic world where people ascribe to different moral systems and therefore abide by different ethical principles. 
Making decisions that align with a group of people with different ethical principles poses a so-called \emph{pluralistic value alignment} problem \cite{gabriel2020artificial}. Indeed, this is the case, for instance, when conducting policy-making decisions that align stakeholders with various ethical principles (e.g., \cite{pigmans2017decision,pigmans2019role}). Indeed, as noted in \cite{Scharfbillig2021ValuesSummary},  
policymakers must consider citizens' values (ethical principles) and identities when developing and communicating policies. We argue that the contributions in this paper make headway in this direction. The tools provided in this paper not only allow a group to reach a consensus but to do so while taking into account the collective ethics of the group. Future work should address the practical use of the tools presented here.

\section*{Acknowledgements}

Research supported by projects:  CI-SUSTAIN (PID2019-104156GB-I00); TAILOR (H2020-952215);  2021 SGR 00754 funded by Generalitat de Catalunya; VAE TED2021-131295B-C31, funded by MCIN/AEI/10.13 039/501100011033 and NextGenerationEU/ PRTR; and VALAWAI (Horizon Europe \#101070930). Funding for open access charge: CRUE-Universitat Politècnica de València.

\bibliography{mybibfile}

\end{document}